\documentstyle[prl,aps,preprint]{revtex}
\newcommand{\ba}{\begin{array}}
\newcommand{\ea}{\end{array}}

\newcommand{\be}{\begin{equation}}
\newcommand{\ee}{\end{equation}}
\newcommand{\bea}{\begin{eqnarray}}
\newcommand{\eea}{\end{eqnarray}}
\newcommand{\beal}{\setcounter{letter}{1} \begin{eqnarray}}
\newcommand{\eeal}{\addtocounter{equation}{1} \end{eqnarray}}

\newcommand{\req}[1]{Eq.(\ref{#1})}

\newcommand{\larrow}{\,\,\,\,\hbox to 30pt{\rightarrowfill}
\,\,\,\,}
\newcommand{\slarrow}{\,\,\,\hbox to 20pt{\rightarrowfill}
\,\,\,}

\newcommand{\spatial}{\int^{\sigma_+}_{\sigma_-}dx}
\begin{document}
\setcounter{page}{0}
\def\footnoterule{\kern-3pt \hrule width\hsize \kern3pt}
\tighten
\title{Exact Physical Black Hole States in Generic 2-D Dilaton Gravity\thanks
{This work is supported in part by funds provided by the U.S.
Department of Energy (D.O.E.).}}
\author{A. Barvinsky}

\address{Theory Department,Lebedev Physics Institue \\
Leninsky Prospekt 53, Moscow 117924, Russia \\
{Email address: {\tt barvin@rq2.fian.msk.su}}
{~}}

\author{G. Kunstatter}
\address{Center for Theoretical Physics and Laboratory for Nuclear Science \\
Massachusetts Institute of Technology, Cambridge, MA 02139 \\
and \\
Physics Department, University of Winnipeg \\
Winnipeg, Man., Canada R3B 2E9\\
{Email address: {\tt gabor@theory.uwinnipeg.ca}}\\[5pt]
{~}}

\date{MIT-CTP-2528,~ hep-th/9606134. {~~~~~}April, 1996}
\maketitle

\thispagestyle{empty}

\begin{abstract}
The quantum mechanics of black holes in generic 2-D dilaton gravity is
considered. The Hamiltonian surface terms are derived for boundary conditions
corresponding to an  eternal black hole with slices on the
interior ending on the horizon bifurcation point.
The quantum Dirac constraints are solved exactly for these boundary conditions
to yield physical eigenstates of the energy operator. The solutions
are obtained in terms of geometrical phase space variables that were
originally
used by Cangemi, Jackiw and Zwiebach in the context of string inspired dilaton
gravity.
The spectrum is continuous in the Lorentzian sector, but in the Euclidean
sector the thermodynamic entropy must
be $2\pi n/G$ where $n$ is an integer. The general class of models considered
contains as special cases string inspired dilaton gravity, Jackiw-Teitelboim
gravity and spherically symmetry gravity.

 \end{abstract}

\vspace*{\fill}
\begin{center}
Submitted to: {\it Physics Letters B}
\end{center}

\pacs{xxxxxx}
\clearpage

\section{Introduction}\medskip
\par
In order to solve fundamental problems associated with Hawking
radiation\cite{hawking} and black hole thermodynamics non-perturbative
quantum gravity is required.
Generic dilaton gravity provides a large class of  models that are
classically and quantum mechanically solvable. They also contain as
special cases physically relevant theories:
Jackiw-Teitelboim gravity(JT)\cite{JT}, the string inspired model(SIG)
used in the seminal paper of CGHS \cite{SIG}, and spherically symmetric
gravity (SSG) \cite{SSG}. A great deal of progress has been made towards
quantizing specific models (\cite{henneaux},\cite{kuchar},\cite{cangemi})
as well as the generic theory (\cite{strobl},\cite{domingo1}). In particular,
Strobl\cite{strobl} has quantized a large class of 2-D gravity theories
(including generic dilaton gravity) with a Poisson-sigma model approach that
generalizes the gauge theoretic formulations of JT and SIG \cite{cangemi}.
In other work \cite{domingo1,DGK}, mass eigenstates have been found in terms
of a metric formulation in a WKB approximation.
\par
The purpose of the present paper is to quantize generic 2-D dilaton gravity
within a metric formulation with black hole boundary conditions. The
specific
boundary conditions we adopt were first considered by Louko and
Whiting\cite{LW} for SSG: we restrict consideration to slices that lie on the
exterior of
an eternal black hole in the generic theory. One end of the slice hits the
bifurcation point along a static slice, while the other
approaches the asymptotic region, also along a static slice. We derive the
Hamiltonian
for these boundary conditions and then quantize the theory.
By adopting canonical variables first used by Cangemi {\it et al}\cite{CJZ}
for SIG we are able to find exact physical eigenstates of the mass operator
and Hamiltonian. We find that while the mass spectrum is continuous in the
Lorentzian sector, when the theory is quantized in the Euclidean sector,
single valuedness of the wave function requires the black hole entropy to
be an integer multiple of 2$\pi$/G. This agrees in form, but not in detail
with recent conjectures by Bekenstein and Mukhanov\cite{bekenstein} concerning
the
quantization
of black hole entropy in Einstein gravity.
\par
Section II reviews some general features of generic dilaton gravity
and derives the Hamiltonian surface terms required in the
case of black hole boundary conditions. In Section III the theory
is quantized in the Lorentzian sector,
using the  phase space variables of Cangemi {\it et al}. Section IV outlines
 the same
calculation for Euclidean black holes and argues that the entropy must
be quantized in this case.
Finally, Section  V closes with conclusions and prospects for future work.
\section{Generic Dilaton Gravity}
\bigskip
\par\noindent
The classical action for generic dilaton gravity in two spacetime
dimensions is
\be
I = {1\over 2 G} \int dt dx \sqrt{-g} \left(\eta R(g)+
   {V(\eta)\over l^2}\right),
\label{eq: action1}
\ee
where $G$ is a (dimensionless) gravitational constant, and $l$ is a fundamental
constant of dimension length.
\req{eq: action1}  is the most general coordinate invariant,
second order action involving a scalar and  metric in
two spacetime dimensions\cite{banks}. It includes as special cases spherically
symmetric
gravity(SSG), for which $V(\eta)= 1/\sqrt{2\eta}$, as well as string inspired
gravity ($V(\eta)=1)$. The most general solution to the field equations in the
generic theory up to spacetime diffeomorphisms can be
written\cite{domingo2}\cite{DGK}:
\bea
ds^2&=&-(j(\eta) - 2GlM)dt^2 + {1\over (j(\eta) - 2GlM)} dx^2,
\nonumber\\
\eta &=& x/l,
\label{eq: general solution}
\eea
where $j(\eta) = \int^\eta_0 d\tilde{\eta} V(\tilde{\eta})$ and the parameter
$M$ will turn out to be the ADM mass. These solutions all have a Killing
 vector, $k^\mu = \epsilon^{\mu\nu}\partial_\nu \eta /\sqrt{-g}$, whose
norm can be written in coordinate invariant form as
\be
|k|^2 = -l^2 |\nabla \eta |^2 = ( 2GlM - j(\eta)).
\label{eq: killing vector}
\ee
In the case that $j(\eta)$ is monotonic, the above solutions describe
Schwarzschild-like black
holes\cite{DGK}.
Note, however, that \req{eq: general solution} does not describe metrics
that are in general asymptotically Minkowskian, nor is the Killing vector
normalized to a constant value at spatial infinity. One can therefore define
a ``physical metric" for which these properties do hold by doing a conformal
reparametrization $\tilde{g}_{\mu\nu}= g_{\mu\nu}/j(\eta)$. The results that
follow apply equally well to the physical metric, but the present
 parametrization simplifies the analysis considerably.
\par
 Details of the corresponding thermodynamics
can be found in \cite{DGK}. The crucial observation for the present purposes
is that event horizons are surfaces of constant $\eta=\eta_0$ for which
$|k|^2=0$. The entropy of the corresponding black hole is proportional to
the value of the dilaton at the horizon, namely
\be
S= {2\pi\over G}\eta_0.
\label{eq: entropy}
\ee
\par
The Hamiltonian formulation for the geometrical theory starts with a
decomposition of the metric:
\be
ds^2 = e^{2\rho}\left(-u^2dt^2 +(dx-vdt)^2\right).
\label{eq: metric decomp}
\ee
A straightforward calculation yields the following form for the action,
up to surface terms:
\be
I = \int dt \spatial \left(\Pi_\rho\dot{\rho}+\Pi_\eta\dot{\eta}
   -u{\cal E}-v{\cal P}\right),
\label{eq: canonical action}
\ee
where $\Pi_\rho$ and $\Pi_\eta$ are momenta canonically conjugate to
$\rho$ and $\eta$, respectively.
%\bea
%\Pi_\rho &=& {1\over G u }\left(\dot{\eta}- v \eta'\right)
 %     \label{eq: pi_rho}\\
%\Pi_\eta &=& {1\over G u }\left(\dot{\rho}-v\rho' - v'\right)
%    \label{eq: pi_eta}
%\eea
and we have defined the Hamiltonian and momentum constraints:
\bea
{\cal E} &=&  -  G \Pi_\rho \Pi_\eta +{1\over 2 G}\left(
   2\eta''-2\rho'\eta'- {V(\eta)\over l^2} e^{2\rho}\right),
\label{eq: ham_constraint}\\
{\cal P} &=& \Pi_\rho' -\Pi_\rho \rho' - \Pi_\eta \eta'.
 \label{eq: mom_constraint}
\eea
\par
It is useful to replace $\cal E$ by the following linear combination
of constraints\footnote{This was first done for JT by Henneaux
\cite{henneaux}.}:
\bea
\tilde{\cal E}\equiv l\,e^{-2\rho}\left(-\eta' {\cal E} + G \Pi_\rho {\cal P}
  \right)= {d{\cal M}\over dx},
\label{eq: tildeE}
\eea
where:
\be
{\cal M}:= {l\over 2 G} \left(e^{-2\rho} ( G^2 \Pi_\rho^2 - (\eta')^2)
  +{j(\eta)\over l^2}\right) = {1\over 2Gl}(|k|^2 + j(\eta)).
\label{eq: define M}
\ee
Clearly $\cal M$ is a constant on the constraint surface. One can verify that
it commutes weakly with the constraints. Furthermore, for
classical solutions \req{eq: general solution}, ${\cal M}= M$ is the ADM mass.
Thus, as discussed in \cite{DGK}, the constant mode of $\cal M$ is a physical
observable, corresponding to the ADM mass of the solution\footnote{It also
corresponds to the Casimir invariant that characterizes solutions in the
Poisson sigma model approach\cite{strobl}.}. It is possible, following
Kuchar\cite{kuchar} to do a canonical transformation in which
$\cal M$ becomes one of the phase space variables, but this will not be
necessary for our purposes. We henceforth call $\cal M$ the mass observable,
 to distinguish it from the total Hamiltonian.
\par
In terms of the new constraint, the canonical Hamiltonian is:
\be
 H_c= \spatial
   \left(-\tilde{u}{\cal M}'+\tilde{v}{\cal P}\right) + H_+-H_-,
\label{eq: canonical hamiltonian1}
\ee
where
\bea
\tilde{u} &=& {u e^{2\rho}\over l\eta'},\\
\tilde{v} &=& v+{uG\Pi_\rho\over \eta'},
\label{eq: tildev}
\eea
$H_+$ and $H_-$ are determined by the requirement that the surface
variations vanish for a given set of boundary conditions.
\par
For concreteness we will choose black hole boundary conditions
similar to those considered
recently by Louko and Whiting\cite{LW} in the case of spherically
symmetric gravity.
 In particular we restrict consideration to slices that are static at both
$\sigma_+$ and $\sigma_-$. $\sigma_+$ will correspond to the asymptotic region
of an eternal black hole, and $\sigma_-$ will be chosen to approach the horizon
bifurcation point. Thus, we require $\tilde{v}_+=0$, so that  only the first
term in \req{eq: canonical hamiltonian1}
contributes a surface term at $\sigma_+$.
Moreover if $\sigma_+$ lies in the asymptotic external region
of the black hole, it can be shown that $\tilde{u}_+\to 1$, from which it
follows immediately that
\be
H_+ = {\cal M}_+ = {\cal M}
\ee
\par
For the surface at $\sigma_-$ to approach the
bifurcation point (i.e. $k^\mu =0$) along a static slice, the following
conditions must be satisfied\cite{LW}:  $u_-=0$, ${v}_- =0$,
$\Pi_\rho(\sigma_-)=0$
and $\eta_-'=0$. With
these boundary conditions, the constraints imply that:
\be
\tilde{u}_- = {u'e^{2\rho}\over l \eta''} = {2 l  N_0\over V(\eta_-)},
\ee
where we have used l'Hopital's rule to get the middle expression. The
final expression was obtained using $\eta''$ from \req{eq: tildeE} with
$\eta'=0$ and $\Pi_\rho=0$.
$N_0:= u'_-$ gives the rate of change of the unit normal to the constant $t$
surfaces at the bifurcation two sphere\cite{LW}. For the on-shell Euclidean
black hole it is proportional to the Hawking temperature.
Using the fact that $|k|_-^2 =0$(cf \req{eq: killing vector}), one has
\be
\delta {\cal M}_- = {V(\eta_-)\over 2G l} \delta \eta_-,
\ee
so that
\be
\tilde{u} \delta M_- = {N_0 \over Gl} \delta \eta_-.
\ee
This can be integrated for fixed $N_0$, so that the total canonical
Hamiltonian is:
\be
H_c= \spatial \left(-\tilde{u}{\cal M'}+\tilde{v}{\cal P}\right)
 + {\cal M}) - {N_0\over 2\pi} S,
\label{eq: canonical Hamiltonian}
\ee
where we have used \req{eq: entropy} to define the
classical thermodynamic entropy:
\be
S= {2\pi\over G}\eta_-.
\ee
Note however that  $\eta_-$  and $ M$ are not independent.
In particular,  \req{eq: define M} requires that on the constraint surface:
\be
j(\eta_-) = 2Gl M.
\ee
\par
\req{eq: canonical Hamiltonian} generalizes Eq.(5.2) of \cite{LW} to the case
of black holes in generic dilaton gravity.
 Since $N_0$ determines the deficit angle for the off-shell black hole, it is
also consistent with the results of Teitelboim\cite{teitelboim} who  showed
for SIG that $N_0$ is conjugate to the entropy.
\smallskip
\section{Quantum Theory}
\par
We now introduce a generalization of the phase space variables first used by
CJZ\cite{CJZ}
in the context of SIG. These variables are also closely related to
Poisson Sigma model variables\cite{strobl}. We
perform the canonical transformation
\be
\{\rho(x),\Pi_\rho(x),\eta(x),\Pi_\eta(x)\} \to
\{\rho^a(x),p_a(x),\eta_-,\theta_-\}\,\, (a=0,1)
\ee
defined by
\bea
\rho^1 &=&e^{-\rho}(\Pi_\rho\sinh \theta - \eta'\cosh \theta),\\
p_1 &=& e^\rho \sinh \theta, \label{eq: p1}\\
\rho^0 &=& e^{-\rho} (\Pi_\rho\cosh \theta - \eta'\sinh \theta),\\
p_0 &=& -e^\rho \cosh \theta, \label{eq: p0}\\
\eta_-&=&\eta(\sigma_-),\,\,\theta_-=\theta(\sigma_-), \label{eq: pi_eta}
\eea
where
\bea
\theta(x) \equiv - \int^{\sigma_+}_x d\tilde{x} \Pi_\eta(\tilde{x})
\label{eq: defn theta}.
\eea
The inverse transformation reads
\bea
e^{2\rho}&=& (p_0^2 - p_1^2)\equiv -p^2, \\
\Pi_\rho &=& -\rho^0 p_0 - \rho^1 p_1, \\
\eta &=& \eta_- + \int^x_{\sigma_-}d\tilde{x} (\rho^0 p_1
+ \rho^1 p_0),
\\
\Pi_\eta &=&-\left[\hbox{arctanh}\left({p_1\over p_0}\right)\right]'
   = -{p_0 p_1' - p_1 p_0' \over p^2}
\label{eq: eta definition}
\eea
and generates a new symplectic form with $\theta_-$ canonically conjugate
to $\eta_-$
\be
\spatial \left(\Pi_\rho \delta{\rho} + \Pi_\eta \delta{\eta}\right)
= \spatial\,p_a\delta{\rho}^a +  \eta_- \delta{\theta}_-.
\label{eq: symplectic form}
\ee
It would seem that the new variables form an overcomplete set because
formally $\theta_-$ is not independent from the momenta $p_a$:
\be
\theta_- = - {\rm arctanh} (p_1/p_0)|_{\sigma_-}.
\label{eq: theta-}
\ee
However, the bifurcation point boundary conditions imply that $\rho^a_-=0$
so that $p_a|_{\sigma_-}$ do not appear in the integral on the right
hand side of \req{eq: symplectic form}. They appear only implicitly
through the variable $\theta_-$. Thus $\theta_-$ is conjugate to
$\eta_-$ and the canonical transformation is non-singular\footnote{Note that
$\theta_-$ is not analytic in terms of new momenta and the nonanalyticity point
does in principle
belong to their range: for  the classical solution in Schwarzschild-like
coordinates (\req{eq: general solution}) ${\rm
exp}[\rho(\sigma_-)]=(j(\sigma_-)-2GlM)^{-1}=\infty$, so that
from (\ref{eq: p1}) and (\ref{eq: p0}) $p_a(\sigma_-)=\infty$
The singularity in the momenta at $\sigma_-$ does not however spoil the
symplectic form because
$p_a(\sigma_-)$ are  multiplied by $\delta\rho^a(\sigma_-)=0$
with $\rho^a(\sigma_-)=0$ at the bifurcation point. Thus the ratio
$(p_1/p_0)(\sigma_-)$ is not determined and should be taken care of by the
extra
variable $\theta_-$ becoming a dynamical momentum conjugate to $\eta_-$.}.
Another argument for including ($\eta_-,\,\theta_-$) as an independent
pair of conjugate variables is that variation of the resulting
action with respect to
$\theta_-$ leads to correct equation of motion for $\eta_-,\,
\dot{\eta}_-=0$.
\par
In terms of the new variables, the mass operator $\cal M$ takes a very simple
form :
\be
{\cal M} ={l\over 2G }\left(-\rho^2 +{ j(\eta)\over l^2}\right),
\ee
where $\rho^2\equiv(\rho^1)^2-(\rho^0)^2$.
The momentum constraint is:
\be
{\cal P}= - \rho^0{}'p_0 - \rho^1{}'p_1.
\ee
\par
One final feature of these variables that is crucial to the quantization is
the fact that $\eta$ and $\rho^2$ have vanishing Poisson bracket.
\par
We now quantize the theory,
in the functional Schrodinger representation, with
wave functionals ${\Psi[\rho^a|\,\theta_-)}$. The notation indicates that
${\Psi[\rho^a|\,\theta_-)}$ is a functional of $\rho^a(x)$, but
an ordinary function of the coordinate $\theta_-$. The momentum
operators are:
\bea
\hat{p}_a &=& -i\hbar {\delta\over \delta \rho^a},\nonumber\\
\hat{\eta}_- &=& -i\hbar {\partial \over \partial \theta_-}.
\label{eq: operators}
\eea
\par
We wish to find physical states that are eigenstates of the Hamiltonian
operator:
\be
\hat{H}= \spatial\left(-\tilde{u}\hat{\cal M}' +\tilde{v}\hat{\cal P}
   +\hat{\cal M}- {N_0\over G} \hat{\eta}_-\right),
\ee
where
\bea
\hat{\cal P} &=& -\rho^0{}'\hat{p}_0 - \rho^1{}'\hat{p}_1, \nonumber\\
\hat{\cal M} &=& {l\over 2G} \left(-\rho^2 + {j(\hat{\eta})\over
l^2}\right),\nonumber\\
\hat{\eta}(x)&=&\hat{\eta}_- +\int^x_{\sigma_-}dx
    (\rho^0 \hat{p}_1 + \rho^1 \hat{p}_0).
\eea
A necessary and
sufficient condition is that the following be satisfied:
\bea
\hat{\cal P}{\Psi[\rho^a|\,\theta_-)} &=&0,\nonumber\\
\hat{\cal M}{\Psi[\rho^a|\,\theta_-)} &=& M {\Psi[\rho^a|\,\theta_-)},
\nonumber\\
\hat{\eta}_- {\Psi[\rho^a|\,\theta_-)} &=& \eta_- {\Psi[\rho^a|\,\theta_-)},
\label{eq: eigenvalue equations}
\eea
where $M$ and $\eta_-$ are (constant) c-numbers.
Inspired by the solution of \cite{CJZ} to SIG, we try the following
ansatz:
\be
{\Psi[\rho^a|\,\theta_-)} = \exp\left({i\over \hbar} \spatial \quad
  \omega(\rho^2)(\rho^0\rho^1{}'- \rho^1\rho^0{}')\right)
  \exp({i\over \hbar}\eta_-\theta_-).
\label{eq: ansatz}
\ee
This is invariant under spatial diffeomorphisms because the integrand is the
product of a scalar and a density. (The $\rho$'s are all
scalars and their derivatives densities).
It can be shown that
\footnote
{We are assuming the boundary condition
at $\sigma_+$ that $(\rho^0_+/\rho^1_+)$ is also fixed. Otherwise the
variation
of ${\Psi[\rho^a|\,\theta_-)}$ would require the
addition of a boundary term to the phase. This boundary condition naturally
arises in the asymptotically static case. From
the classical solution (\ref{eq: general solution}) and $\theta_+=0$
it follows that $(\rho^0/\rho^1)_+=-\Pi_\rho/\eta'|_+=0$.
}:
\be
\hat{\eta} {\Psi[\rho^a|\,\theta_-)} = (\eta_-
+\omega\rho^2){\Psi[\rho^a|\,\theta_-)},
\ee
where we have used
black hole boundary condition $\rho^2_-=0$.
Now we use the fact that  $[\hat{\eta}, \rho^2]=0$ (there is no anomaly in
this commutator because $\rho^2=(\rho^1)^2-(\rho^0)^2$ is a 2D indefinite
quadratic form and anomalies coming from $\rho^1$ and $\rho^0$ sectors
cancel out \cite{CJZ}): the action of $j(\hat{\eta})$
on ${\Psi[\rho^a|\,\theta_-)}$, for any $j(\eta)$ that has a Taylor expansion,
is simply:
\be
j(\hat{\eta}){\Psi[\rho^a|\theta_-)} = j(\eta_-
+\omega\rho^2){\Psi[\rho^a|\theta_-)}.
\ee
Thus, the state will be an eigenstate of the ADM mass operator
$\hat{\cal M}$ with eigenvalue $M$ if:
\be
j(\eta_- +\omega \rho^2) - l^2 \rho^2 = 2Gl M.
\label{eq: eigenvalue equation}
\ee
The eigenvalue $\eta_-$ is  fixed by \req{eq: eigenvalue equation}
and the boundary
condition $\rho^2_-=0$ to be:
\be
\eta_- = j^{-1}(2GlM).
\ee
The solution for $\omega$ is therefore:
\be
\omega(\rho^2) = {1\over \rho^2}\left(
    j^{-1}(2GlM+l^2\rho^2) - j^{-1}(2GlM)\right).
\label{eq: omega solution}
\ee
\par
 With $\omega$ and $\eta_-$ given
above the wave functional \req{eq: ansatz}
represents a physical eigenstate of both the ADM mass operator
$\hat{\cal M}$ and the entropy operator $\hat{\eta_-}$. For
a given mass $M$, the
eigenvalue of the entropy is fixed to be its classical value,
and the corresponding eigenvalue of the Hamiltonian is:
\be
E = M - {N_0\over G} j^{-1}(2GlM).
\ee
\par
Note that for SIG, $j(\eta) = \eta$, which gives
\be
\omega = l^2
\ee
in agreement with the result of \cite{CJZ}.
\footnote{CGZ obtained only two states (as opposed to the continuous
spectrum found here) because  they imposed stronger boundary
conditions. We are grateful to E. Benedict for conversations on
this point.}
\par
\smallskip
\section{Euclidean Black Holes}
\par
We now briefly describe the result of quantizing the black holes in
the Euclidean sector. The Euclidean version of the theory originates
from the Wick rotation to imaginary (Euclidean) time:
   $t= -i t_E$,
   $v= i v_E$
(the latter redefines the shift to absorb an extra factor of $i$ in the metric
so that everything is real) and  boils down to imaginary values of
momenta
   $\Pi_\rho= -i \Pi_\rho^E$,
   $\Pi_\eta= -i \Pi_\eta^E$,
so that the mass operator in terms of the Euclidean momentum acquires
a negative kinetic term
\be
M= -\frac l{2G}e^{-2\rho}((\eta')^2 +\Pi_\rho^2) + \frac {j(\eta)}{2Gl}.
\ee
{}From (\ref{eq: defn theta}) it is obvious that $\theta$ is also subject
to Wick rotation
$\theta=i\theta_E$ whence it follows that in the canonical transformation
to CJZ variables the hyperbolic sine and cosine go over into trigonometric
functions of an angular variable $\theta_E$ and, moreover, $\rho^0$
and $p_1$ become imaginary: $\rho^0=i\rho^0_E,\,p_1=i p_1^E$. Thus the
coordinates $\rho^a$ under Euclideanization transform as the Wick rotation of a
``timelike" coordinate, emphasizing their
geometric nature as embedding variables in the Lorentzian
and Euclidean 2D spacetimes. For Euclidean variables the CJZ canonical
transformation
\bea
 e^{2\rho} &=& (p^E_0)^2 + (p_1^E)^2, \nonumber\\
\Pi_\rho^E &=& - (\rho^0_E p_0^E + \rho^1_E p_1^E), \nonumber\\
\eta (x)&=& \eta_- - \int^x_{\sigma_-} d\tilde{x} (\rho^0_E p_1^E- \rho^1_E
p_0^E),
\nonumber\\
\Pi_\eta^E &=& -\left [\arctan (p_1^E/p_0^E) \right]'
\eea
generates the same symplectic form (\ref{eq: symplectic form}) with
$\theta_- = -\arctan(p_1^E/p_0^E)|_-$ and yields a mass operator
$2G{\cal M} =  -l\,\rho_E^2 + j(\eta)/l,\,\rho_E^2\equiv
(\rho_E^0)^2+(\rho_E^1)^2$
for which one can find exact quantum eigenstates precisely as before.
Now however the angular variable $\theta_-$ determines the orientation of
the vector $p^E_a$ in the 2D Euclidean momentum plane, and it is natural to
demand that the physical state is periodic in this variable. Therefore,
single valuedness of the factor $\exp(i\eta_-\theta_-/\hbar)$ in the
wavefunction requires the eigenvalue $\eta_-$ of the dilaton operator
to be an integer in $\hbar$ units. The black hole entropy is therefore
quantized: $S(M) = 2\pi\hbar n/G$. Remarkably this gives the same spectrum
for Euclidean black holes as reduced quantization\cite{bk}.

\smallskip
\section{Conclusions}
We have quantized generic dilaton gravity theory exactly in terms
of geometrical variables and derived physical black hole states that are
eigenstates of the Hamiltonian, mass operator and entropy operator.
In previous work\cite{domingo2},  the constraints were first linearized
in the momenta
and then solved exactly at the quantum level. The resulting states were only
eigenstates of the
mass operator in the WKB approximation.  Louis-Martinez\cite{domingo3}
has been able to show that the  states presented here are
equivalent to those presented in \cite{domingo2}, thus generalizing
Benedict's proof\cite{benedict} for SIG. In retrospect this is not surprising.
In the present
parametrization we have found that the WKB approximation is exact. To WKB
order it is always possible to find the transformation relating states
derived in two different parametrizations. What the
present analysis shows is that
there exists a factor ordering for the mass operator for which the WKB states
found in \cite{domingo2} are also exact.
\par
It is clearly of interest to explore the properties of these wave functionals
in more detail and, in particular, to understand the meaning of the
quantization of the theory in the Euclidean sector. It should be emphasized
that the procedure we used was not the analytic continuation of the
Lorentzian quantum state to the Euclidean section of phase space, but
first the Euclideanization of the theory with its subsequent quantization.
An important difference of this procedure from the usual transition to
Euclidean spacetime in nongravitational theories is a complexification
of the phase space resulting here in a Wick rotation of the ``timelike"
CJZ coordinate $\rho^0$. With the other choice of variables for
quantization this complexification might look very involved and
geometrically vague, while in CJZ parameterization this Euclideanization
has a geometrically covariant form. Clearly, Euclidean 2D dilaton
gravity should be related to the physical theory in Lorentzian spacetime
just like in ordinary theories: it describes  thermodynamical
and underbarrier penetration properties or serves as a powerful
calculational tool for quantum transition amplitudes. Understanding this
relation might give an insight into the quantized nature of the black
hole mass and entropy and explain the statistical mechanical origin of the
latter. Finally, given
the utility of the CJZ variables in the pure gravity case, it is likely that
the quantization of the generic theory with matter would also be possible
using these techniques. This is currently under investigation.

\section{Acknowledgements}
\par
GK is grateful to E. Benedict,
V. Frolov, J. Gegenberg and D. Louis-Martinez for helpful
discussions and to the  CTP at MIT for
its hospitality during the completion of this work. AB thanks
for hospitality and financial support the
Winnipeg Institute for Theoretical Physics and the University of
Winnipeg where this work was  started . This work
was supported in part by the Natural Sciences and Engineering
Research
Council of Canada. The work of AB was also
supported by the the Russian Foundation for Basic Research under
Grants 96-02-16287 and 96-02-16295, the European
Community Grant INTAS-93-493 and the Russian Research Project
"Cosmomicrophysics".

  \par

\end{document}